\global\def\draftcontrol{0}

   \def\versionno{ strings->BH -- draft   }

\catcode`\@=11

\expandafter\ifx\csname draftcontrol\endcsname\relax\global\def\draftcontrol{0}
\fi

{\count255=\time\divide\count255 by 60
\xdef\hourmin{\number\count255}
\multiply\count255 by-60\advance\count255 by\time
\xdef\hourmin{\hourmin:\ifnum\count255<10 0\fi\the\count255}}
\def\draftdate{\number\month/\number\day/\number\year\ \ \ \hourmin }

\newcommand\makepapertitle{\par
  \begingroup
    \renewcommand\thefootnote{\@fnsymbol\c@footnote}%
    \def\@makefnmark{\rlap{\@textsuperscript{\normalfont\@thefnmark}}}%
    \long\def\@makefntext##1{\parindent 1em\noindent
            \hb@xt@1.8em{%
                \hss\@textsuperscript{\normalfont\@thefnmark}}##1}%
     \newpage
     \global\@topnum\z@   
     \@makepapertitle
     \thispagestyle{empty}\@thanks
  \endgroup
  \setcounter{footnote}{0}%
  \global\let\thanks\relax
  \global\let\makepapertitle\relax
  \global\let\@makepapertitle\relax
  \global\let\@thanks\@empty
  \global\let\@author\@empty
  \global\let\@date\@empty
  \global\let\@title\@empty
  \global\let\title\relax
  \global\let\author\relax
  \global\let\date\relax
  \global\let\and\relax
  \def\version{\let\version\@version\@gobble}
}
\def\@makepapertitle{%
  \newpage
   \ifnum\draftcontrol=1 {}
   \version\versionno
   \vskip 3em%
   \else
   \hfill\hbox to 3cm {\parbox{4cm}{\@pubnum}\hss}%
   \vskip 3em%
   \fi
   \begin{center}%
   \let \footnote \thanks
     {\LARGE {\@title}}%
     \vskip 1.5em%
     {\normalsize
       \lineskip .5em%
       \begin{tabular}[t]{c}%
         \@author
       \end{tabular}\par}%
     \vskip 1.5em%
     {\@bstract}%
     \end{center}%
     \vskip 1.5em 
     \@date%
   \par
}

\gdef\@pubnum{}
\def\pubnum#1{%
  \gdef\@pubnum{#1}}

\gdef\@bstract{}
\def\Abstract#1{%
  \gdef\@bstract{%
   \parbox{\textwidth-0pc}{%
   \centerline{\bf Abstract}\penalty1000%
\noindent
\renewcommand\baselinestretch{1.0}%
{#1}}}
}

\def\ps@paper{\let\@mkboth\@gobbletwo%
     \ifnum\draftcontrol=1
        \def\@oddfoot{\hbox to \textwidth{\tiny \versionno \hfil\tiny\draftdate}%
        \hskip -\textwidth \hbox to \textwidth{\hfil\rm\thepage\hfil}}%
     \else\def\@oddfoot{\hbox to \textwidth{\hfil\rm\thepage\hfil}}
     \fi
     \let\@evenfoot\@oddfoot
}

\def\body{\clearpage
          \pagestyle{paper}
        }

\def\@version#1{\ifnum\draftcontrol=1
\typeout{}\typeout{#1}\typeout{}
\vskip3mm\centerline{\hbox{\fbox{\normalsize{\tt DRAFT -- #1 -- }
                   {\draftdate}}}}\vskip3mm
\fi}
\let\version\@version
\long\def\eqlabel#1{\ifnum\draftcontrol=1
                    \tag@false  
                    \tag*{(\theequation) \hbox to -0.2cm{\hspace{0cm}\small{#1}\hss}}
                    \refstepcounter{equation} 
                    \edef\@currentlabel{\theequation}
                    \ltx@label{#1}          
                    \else
                    \label{#1}
                    \fi
                    }
\let\st@bibitem\@bibitem
\let\st@lbibitem\@lbibitem
\ifnum\draftcontrol=1
  \def\@bibitem#1{%
    \st@bibitem{#1}\a@@label{#1}\ignorespaces}
  \def\@lbibitem[#1]#2{%
    \st@lbibitem[#1]{#2}\a@@label{#2}\ignorespaces}
  \def\a@@label#1{%
    \gdef\a@lab{\smash{\normalfont\small#1}}
    \ifvmode
      \if@inlabel
        \global\setbox\@labels\hbox{%
          \llap{\a@lab\let\a@lab\relax
                \kern\@totalleftmargin\kern\marginparsep}%
          \box\@labels}%
      \fi
    \fi}
\fi

\documentclass[12pt,letterpaper]{article}

\usepackage{amsmath,amssymb,array,calc,rotating,epsfig,psfrag}
\usepackage[nosort]{cite}

\ifnum\draftcontrol=1
\fi

\renewcommand\baselinestretch{1.25}
\renewcommand\section{\@startsection {section}{1}{\z@}%
                                   {-3.5ex \@plus -1ex \@minus -.2ex}%
                                   {2.3ex \@plus.2ex}%
                                   {\normalfont\large\bfseries}}
\renewcommand\subsection{\@startsection{subsection}{2}{\z@}%
                                   {-3.25ex\@plus -1ex \@minus -.2ex}%
                                   {1.5ex \@plus .2ex}%
                                   {\normalfont\normalsize\bfseries}}
\renewcommand\subsubsection{\@startsection{subsubsection}{3}{\z@}%
                                   {-3.25ex\@plus -1ex \@minus -.2ex}%
                                   {1.5ex \@plus .2ex}%
                                   {\normalfont\normalsize\it}}
\renewcommand\paragraph{\@startsection{paragraph}{4}{\z@}%
                                   {-3.25ex\@plus -1ex \@minus -.2ex}%
                                   {1.5ex \@plus .2ex}%
                                   {\normalfont\normalsize\bf}}

\def\ie{{\it i.e.}}

\def\revise#1       {\raisebox{-0em}{\rule{3pt}{1em}}%
                     \marginpar{\raisebox{.5em}{\vrule width3pt\
                     \vrule width0pt height 0pt depth0.5em
                     \hbox to 0cm{\hspace{0cm}{%
                     \parbox[t]{4em}{\raggedright\footnotesize{#1}}}\hss}}}}

\newcommand\nxt[1]  {\\\fnxt#1}

\def\cala         {{\cal A}}

\def\call         {{\cal L}}
\def\calm         {{\cal M}}
\def\caln         {{\cal N}}

\def\calv         {{\cal V}}

\def\reals        {{\mathbb R}}

\def\del          {\partial}

\def\ee           {{\rm e}}

\def\tr           {\mathop{\rm Tr}}

\def\Im           {{\rm Im\hskip0.1em}}

\def\sqr#1#2{{\vcenter{\vbox{\hrule height.#2pt  
 \hbox{\vrule width.#2pt height#1pt \kern#1pt
 \vrule width.#2pt}\hrule height.#2pt}}}}

\newcommand{\ft}[2]{{\textstyle{\frac{#1}{#2}}}}
\def\jsquare{\mathop{\mathchoice{\sqr{8}{32}}{\sqr{9}{12}}
{\sqr{6.3}{9}}{\sqr{4.5}{9}}}}

\def\a{\alpha}
\def\b{\beta}
\def\r{\rho}
\def\rp{r_+}
\def\tq{\tilde{q}}
\def\om{\Omega}
\def\la{\lambda}

\def\be{\begin{equation}}
\def\ee{\end{equation}}

\def\m{\mu}
\def\g{\gamma}
\def\l{\lambda}

\catcode`\@=12

\begin{document}


\title{Hagedorn vs.~Hawking-Page transition \\
in string theory}

\pubnum{%
MCTP-03-27 \\
hep-th/0305179}
\date{May 2003}

\author{
Alex Buchel and Leopoldo A. Pando Zayas\\[0.4cm]
\it Michigan Center for Theoretical Physics \\
\it Randall Laboratory of Physics, The University of Michigan \\
\it Ann Arbor, MI 48109-1120 \\[0.2cm]
}

\Abstract{We study the supergravity dual to the 
confinement/deconfinement phase transition 
for the $\caln=4$ $SU(N)$ SYM on $\reals\times S^3$ 
with a  chemical potential conjugate to a 
$U(1)\subset SO(6)_R$ charge. The appropriate supergravity 
system is a single charge black hole in  
$D=5$ $\caln=8$ gauged supergravity. 
Application of the gauge/string theory 
holographic renormalization approach leads to 
new expressions for the black hole ADM mass and
its generalized free energy. 
We comment on the relation of this phase transition 
to the Hagedorn transition for strings in the maximally supersymmetric
plane wave background with null RR five form field strength.    
}


\makepapertitle

\body

\version\versionno

\section{Introduction}
The gauge/string theory correspondence \cite{m9711,a9905}
relates $\caln=4$ $SU(N)$ supersymmetric Yang-Mills  (SYM)
theory to type IIB string theory on the $AdS_5\times S^5$ background. 
One of the most interesting aspects of this correspondence is that
it allows for comparison between quantities that are not protected by
symmetries. One class of these non-BPS quantities naturally arises by
considering each side of the correspondence at finite temperature. 
For the $AdS_5$ in global coordinates the background 
geometry is dual to the SYM theory in $\reals \times 
S^3$ which opens the possibility of phase transitions. These 
phase transitions are among the typical dynamical processes one expects to be
able to address within the correspondence. 
There are various arguments supporting the existence of a
confining/deconfinement phase transition for the gauge theory in the
strict  $N\to \infty$ limit. The nature 
of the transition, however, is very different at small and large 
't Hooft coupling. Namely, for $\la\gg 1$ 
this  confinement/deconfinement phase
transition can be identified \cite{w97}  with the 
Hawking-Page  (HP) phase transition in an AdS
background \cite{hp}. On the other hand, 
at $\la=0$, it was shown in \cite{sun} 
that the density of gauge invariant states of $\caln=4$
SYM on $S^3$ exhibits a Hagedorn behavior. Based on these arguments, 
Polyakov \cite{polyakov} has suggested an 
interpolating formula for the density of
states: $d(\Delta)\sim e^{b(\la)\Delta}$, with $b(\la)\sim {\rm constant}$
for small $\la$ and $b(\la)\sim \la^{-1/4}$ for large $\la$. 

A very peculiar behavior is expected for the free energy of the gauge 
theory. The $\l=0$ calculation \cite{sun} suggests generically a 
 partition function near the Hagedorn transition of the form
$$
Z(x) \approx \frac{1}{(x_H-x)\xi'(x_H)},
$$
where $x=e^{-1/T}$. From here we see that the free energy is divergent:
$Z=e^{-\beta F}$,
$$
F=\frac{1}{ \beta}\ln(x_H-x)\xi'(x_H).
$$
On the other hand, at strong coupling we expect the free energy to be
finite and given by the appropriate interpretation of the free energy of
the gravity solution.  The natural conjecture is 
that for a generic value of $\la$ the string theory partition function in $AdS_5\times S^5$ 
would have a Hagedorn  temperature precisely equal to the 
Hawking-Page critical temperature. Clearly, 
to substantiate this claim one needs to know  the quantization 
of strings in $AdS_5\times S^5$ with background Ramond-Ramond 
fields, which is currently not understood\footnote{An interesting 
result which might help solve this problem was reported 
recently in \cite{bpr}.}.

While  quantization of strings in the full $AdS_5\times S^5$ background
is not understood completely, string theory in a particular 
Penrose-G\"uven limit is exactly soluble in the light-cone
\cite{metsaev}. The precise dictionary between  string theory
quantities and a particular large R-charge sector of $\caln=4$ SYM 
was formulated in \cite{bmn}. Moreover, the statistical mechanics 
of these string theories has been studied both 
in the canonical \cite{pv,gost} 
and the grand canonical  \cite{gss,blt} ensembles\footnote{
Finite temperature string theory on various pp-wave 
backgrounds has been also discussed in \cite{sug}.}.  
It was found, that, much like in the case of flat  
space, strings in plane waves have a Hagedorn temperature. 
In the case  of the grand canonical ensemble 
(where in addition to the temperature one introduces a 
chemical potential conjugate to the $U(1)\subset SO(6)_R$ 
charge) the free energy  was shown to be finite
near the Hagedorn temperature \cite{blt}: $F\sim \sqrt{\b-\b_H}$. This suggests the possibility
of a phase transition. However, the specific heat is negative and diverges: $c_V\sim
(\b-\b_H)^{-3/2}$, obscuring the nature of the transition.

Following the relation 
between the Hagedorn transition and Hawking-Page 
phase transition outlined above, it is natural to 
ask whether the Hagedorn physics of strings in PP 
wave background is related to the physics of the 
Hawking-Page phase transition for the 
black holes in global $AdS_5$ that carry large $U(1)_J\subset SO(6)_R$
R-charge. The study of this connection is the main 
motivation of this paper. 

In the next section we discuss the thermodynamics of a single 
charge black holes in global $AdS_5$ geometry. 
The explicit solutions were constructed previously in
\cite{bcs} and some aspects of the  thermodynamics were studied 
in   \cite{cg}. We present an alternative 
computation of the charged  
black hole thermodynamic properties, which utilizes the 
holographic renormalization approach of \cite{bk}.
As we explain, this leads to a different 
expression for the ADM mass than the one used in,
say,  \cite{cg}. In section 3 we  review the relevant 
aspects  of the thermodynamics of strings in PP background in the 
grand canonical ensemble \cite{gss}.  
Unfortunately, we find that  the regime where black holes have 
a large $J$-charge corresponds to  very different 
values of the conjugate chemical potential $\mu_J$, 
from  the one used to define the 
'corresponding' PP-wave string grand canonical partition
function.

\section{A single charge black holes in five dimensional 
gauged supergravity}

In this section we discuss the construction (and the thermodynamics) of 
a single charge black hole solution in $D=5$ $\caln=8$ gauged supergravity. 
The asymptotic background is the global $AdS_5$, and the 
black hole would carry a $U(1)\subset SO(6)_R$ electric charge. 
This solution was originally found in \cite{bcs} as a special case of 
the STU-model. The thermodynamic properties of these black
holes were studied previously in \cite{cg}. 
We present a new computation for the thermodynamics, 
which gives different results for the ADM mass (and 
appropriately the Euclidean action) than the one used in 
\cite{bcs,cg}, and in many subsequent papers.

\subsection{The black hole geometry} 
The black hole geometry can be obtained \cite{bcs} as a solution 
of the $D=5$ $\caln=8$ gauged supergravity. 
The relevant effective five-dimensional action 
is 
\begin{equation}
\begin{split}
S_5&=\frac {1}{4\pi G_5}\int_{\calm_5}d^5\xi\ \sqrt{-g}\call
\\
&=\frac {1}{4\pi G_5}\int_{\calm_5}d^5\xi\ \sqrt{-g}
\biggl(\ft 14 R+\ft 12 g^2 \calv-\ft {1}{16} H^{4/3} F^2-\ft {1}{12} 
H^{-2} \left(
\del H\right)^2\biggr)\,, 
\end{split}
\eqlabel{5action}
\end{equation}  
where $g$ is the gauge coupling, $R$ is the scalar curvature,
$F_{\mu\nu}$ is a $U(1)$ field-strength tensor, and $\calv$ 
is the $H$ scalar potential
\begin{equation}
\calv= 2 H^{2/3}+4 H^{-1/3}\,.
\eqlabel{pot}
\end{equation}
From \eqref{5action} we find the following equations of motion
\begin{equation}
\begin{split}
\jsquare H =& H^{-1} \left(\del H\right)^2
+\ft 12 H^{7/3} F^2-3 g^2 H^2\frac{\del \calv}{\del H}\,,\\
0=&\del_\mu\left(\sqrt{-g} H^{4/3} F^{\mu\nu}\right)\,,\\
R_{\mu\nu}=&\ft 12 H^{4/3} F_{\mu\gamma} F_{\nu}\ ^\gamma
+\ft 13 H^{-2} \del_\mu H \del_\nu H-g_{\mu\nu} \left[
\ft 23 g^2 \calv+\ft {1}{12} H^{4/3} F^2\right]\,.
\end{split}
\eqlabel{eom}
\end{equation}
We take the following ansatz for the charged black hole metric 
\begin{equation}
ds_5^2=-e^{-2A/3} f dt^2+e^{A/3 }\left(f^{-1} dr^2 +
r^2 \left(dS^3\right)^2\right)\,,
\eqlabel{metric5} 
\end{equation}
where $A,f$ are functions of the radial coordinate $r$ only, and 
$\left(dS^3\right)^2$ is the round metric on the unit radius 
$S^3$. Additionally, we take $H\equiv H(r)$, and the only nonvanishing 
component of the gauge potential, $F= dA$,  
$A_t\equiv A_t(r)$.  
With this ansatz, using the equations of motion \eqref{eom},
we can rewrite the gravitational Lagrangian in 
\eqref{5action} as a total derivative
\begin{equation}
\sqrt{-g}\biggl(\ft 14 R+\ft 12 g^2 \calv
-\ft {1}{16} H^{4/3} F^2-\ft {1}{12} H^{-2} \left(
\del H\right)^2\biggr)=-\bigg[\ft {1}{12} A' f r^3+\ft 12
r^2 (f-1)\bigg]'\,,
\eqlabel{rel}
\end{equation}
where primes denote derivatives with respect to $r$.
 
Omitting the computational details, 
the relevant two-parameter family $\{\mu,\r\}$ 
of solutions of \eqref{eom} is 
\begin{equation}
\begin{split}
&e^{A}=H\,,\qquad 
f=1-\ft {\mu}{r^2}+g^2 r^2 H\,,\qquad H=1+\ft {q}{r^2}\,,
\qquad A_t=\ft{\tq}{r^2+q}\,,
\end{split}
\eqlabel{sol}
\end{equation}
where we introduced 
\begin{equation}
q=\mu \sinh^2\r\,,\qquad \tq=\mu\sinh\r\cosh\r\,.
\eqlabel{qqt}
\end{equation}
In what follows it will be important that $\tq$ are the 
physical charges (\ie, 
the conserved charges  to which Gauss's law applies).
Note a useful relation 
\begin{equation}
\tq^2=q (q+\rp^2)(1+g^2 \rp^2)\,.
\eqlabel{us1}
\end{equation}

\subsection{The thermodynamics}
Given the explicit single charge black solution \eqref{sol}, 
it is straightforward to extract its  thermodynamics.
The outer black hole horizon is at $\rp$, the 
largest non-negative zero of the function $f$,
\begin{equation}
f(\rp)=0\,.
\eqlabel{oh}
\end{equation} 
The inverse Hawking temperature $\b\equiv \ft {1}{T_H}$, and the 
Bekenstein-Hawking entropy $S_{BH}$ are given by 
\begin{equation}
\begin{split}
\b&=\frac {1}{T_H}=2\pi\ \frac{(\rp^2+q)^{1/2}}{1+g^2 q +2 g^2 \rp^2}\,,\\
S_{BH}&=\frac{\cala_{horizon}}{4 G_5}=\frac{\pi^2}{2 G_5}\ \rp^2 
\left(\rp^2+q\right)^{1/2}\,.
\end{split}
\eqlabel{ts}
\end{equation} 
The computation of the generalized free energy 
$\om\sim \ft 1\b I_E$  (determined 
from the properly regularized Euclidean gravitational action $I_E$) 
and the ADM mass $M$ is  slightly more subtle and we discuss this in 
some details. In \cite{bcs,cg}, the ADM mass is taken to be
\begin{equation}
M(\mu,q)\sim \ft 32 \mu +q\,.
\eqlabel{wrongmass}
\end{equation}  
In particular, \eqref{wrongmass} vanishes for $\mu=q=0$, 
which implies that the mass of global $AdS_5$ is assumed to 
be zero. But the latter statement contradicts the gauge/string
theory correspondence \cite{m9711,a9905}: it was shown 
in  \cite{bk} that $M_{AdS_5}> 0$, moreover its precise value 
exactly coincides with the (positive) dual gauge theory Casimir energy.
It is instructive to see what goes wrong with the prescription 
for computing the ADM mass  used in  \cite{bcs}.
In \cite{bcs}, following the proposal of \cite{hm}, the 
ADM mass of the geometry \eqref{metric5} was defined as the following 
surface integral at radial infinity
\begin{equation}
M=-\frac {1}{8\pi G_5}\int_{S^3}\ N\left(K-K_0\right)\,,
\eqlabel{hmf}
\end{equation}
where $N$ is the norm of the time-like Killing vector 
and $K$ is the extrinsic curvature, dependent on the black 
hole parameters $\{\mu,q\}$; finally $K_0$ is taken to be the 
corresponding extrinsic curvature of the global $AdS_5$ geometry.
Literally following this prescription for the single charge black
of interest here does not give \eqref{wrongmass}, rather 
we find $M\sim g^2 q\ r^2\to \infty$. The technical reason for this is 
that the 5D gauge  fields of the black hole solution
modify the first subleading (as $r\to \infty$) correction
of $f$ in \eqref{sol}, that can not be subtracted by comparing 
with the 'uncharged' global $AdS_5$ geometry. Now, 
the finite mass term results from the second subleading term in 
$f$, and thus the subtraction  \eqref{hmf} necessarily 
would give diverging result\footnote{The application of the procedure
of \cite{hm} to the recently 
studied black hole solution \cite{bl} would also 
give diverging result for the ADM mass. Again, the problem appear to 
be due to the additional matter fields compared to 
the extremal geometry.}.   

In the rest of this section we present the modified prescription 
for computing $I_E, M$,  that, first of all, gives  finite 
results for the above quantities. Additionally, 
in the limit of vanishing charge and the nonextremality 
parameter we recover the global $AdS$ mass of \cite{bk}.    
Our prescription relies on the Maldacena proposal for the 
existence of the dual (local) four dimensional quantum field 
theory for the black hole 
geometry \eqref{metric5}. 
In a sense, this is a simple application of the renormalization 
ideas of \cite{bk}, which was originally implemented in \cite{cj2}
for the non-extremal black holes in AdS, charged under the 
diagonal $U(1)_{diag}\subset U(1)^3\subset SO(6)_R$.   
Unlike the single charge black hole solutions of interest here, 
for the $U(1)_{diag}$ charged black holes, the ADM mass can also 
be computed \cite{cj1} by subtracting the global $AdS_5$ geometry 
as a reference, \eqref{hmf}. Both computations \cite{cj1,cj2}
lead to the same results for the black hole mass and the regularized 
Euclidean gravitational action.

We begin by summarizing our results:
\begin{equation}
\begin{split}
I_E&=\frac {\b\pi}{G_5}\biggl(-\ft 18\mu -\ft {1}{12}q^2 g^2
+\ft {3}{32}g^{-2}+\ft 14 \rp^2\biggr)\,,\\
M&=\frac{\pi}{G_5}\biggl(\ft 38 \mu +\ft 14 q-\ft{1}{12} q^2 g^2
+ \ft {3}{32}g^{-2}
\biggr)\,.
\end{split}
\eqlabel{results}
\end{equation}
Notice that using \eqref{ts} and \eqref{results} we find the 
expected thermodynamics relation\footnote{Black holes 
with 'hair' for which similar relation can be proved 
are discussed in \cite{gtv,bl}.
}
\begin{equation}
I_E=\b \left(M-{\mu_{\tq}}\ \tq\right)-S_{BH} \,,
\eqlabel{thermrel}
\end{equation}  
where $\mu_{\tq}$ is the chemical potential conjugate to the 
physical black hole charge $\tq$, related to the gauge potential
$A_t$ \eqref{sol} at the horizon, $r=\rp$
\begin{equation}
\mu_{\tq}=\frac{\ vol(S^3)}{8\pi G_5}\ A_t(r)\bigg|_{r=\rp}
=\frac{ \pi}{4 G_5}\ \frac{\tq}{\rp^2+q}\,.
\eqlabel{mutq}
\end{equation}
To obtain \eqref{thermrel} we used \eqref{us1}.
In the limit of vanishing charge we find from \eqref{results} 
\begin{equation}
M\bigg|_{q=0}=\frac {3\pi}{32g^2 G_5}+\frac{3\pi \mu}{8G_5} \,,
\eqlabel{zeroq} 
\end{equation}   
which with identification $\mu\equiv r^2_0,\ g\equiv\ft 1\ell$
is precisely the result obtained in \cite{bk} for the 
Schwarzschild black hole in global $AdS_5$.

Let us compute the renormalized (in the sense of \cite{bk}) 
Euclidean gravitational action $I_E$ of \eqref{5action}.
First, we regularize \eqref{5action} by introducing 
the boundary $\del \calm_5$ at fixed (large) $r$ with the 
unit orthonormal space-like vector $n^\mu\propto \delta^{\mu}_r$
\begin{equation}
\begin{split}
S_5^r&=\frac {1}{4\pi G_5}\int_{\rp}^r dr \int_{\del\calm_5}d^4\xi
\sqrt{g_E}\call_E=-
\frac {1}{4\pi G_5}\int_{\rp}^r dr \int_{\del\calm_5}d^4\xi
\sqrt{-g}\call\\
&=\frac {1}{4\pi G_5}\int_{\rp}^r dr
\bigg[\ft {1}{12} A' f r^3+\ft 12
r^2 (f-1)\bigg]'\int_{\del\calm_5}d^4\xi\\
&=\frac{\b\pi}{2G_5}\bigg[\ft {1}{12} A' f r^3+\ft 12
r^2 (f-1)\bigg]\bigg|_{\rp}^r\,,
\end{split}
\eqlabel{rega}
\end{equation}
where the subscript $\ _E$ represents that all the quantities are to be 
computed in Euclidean signature. We used \eqref{rel} to 
obtain the second identity in \eqref{rega}. 
As usual, to have a well-defined variational  problem in the presence
of a  boundary requires the inclusion of the Gibbons-Hawking 
$S_{GH}$ term
\begin{equation}
\begin{split}
S_{GH}&=-\frac{1}{8\pi G_5}\int_{\del\calm_5} d^4\xi\sqrt{h_E} \nabla_\mu n^\mu\\
&=\frac{\b\pi}{2G_5}
\bigg[-\ft {1}{12} A' f r^3-\ft 14
r^3 f'-\ft 32 r^2 f\bigg]\,,
\end{split}
\eqlabel{sgh}
\end{equation}      
where $h_{\mu\nu}$ is the induced metric on $\del\calm_5$
\begin{equation}
h_{\mu\nu}\equiv g_{\mu\nu}-n_\mu n_\nu\,.
\eqlabel{hmet}
\end{equation}
Finally, as in \cite{bk}, we supplement the combined 
regularized action $\left(S_5^r+S_{GH}\right)$ by the appropriate boundary 
counterterms constructed from the local\footnote{The locality 
condition is very important and it follows from the locality 
of the dual gauge theory. In our case this dual gauge theory   
is the  $\caln=4$ SYM in the deconfined phase with a 
given chemical potential conjugate to the $R$-charge. 
Example where locality condition does not hold 
will be discussed elsewhere \cite{ab}.
} 
metric invariants on the boundary 
$\del\calm_5$ 
\begin{equation}
\begin{split}
S^{counter}&=\frac{1}{4\pi G_5}\int_{\del\calm_5}d^4\xi \left(
\a_1\sqrt{h_E}+\a_2 R_4\sqrt{h_E}\right)\\
&=\frac{\b\pi}{2 G_5}\left[\a_1\
e^{A/6} f^{1/2} r^3+6 \a_2\
r f^{1/2}e^{-A/6}
\right]\,,
\end{split}
\eqlabel{scount}
\end{equation}
where $R_4\equiv R_4(h_E)$ is the  Ricci scalar constructed from $h_{\mu\nu}$,
and $\a_i$ are coefficients that can depend only on the 
curvature  of the asymptotic $AdS_5$ geometry\footnote{
For this reason the values of $\a_i$ must be the same as the
corresponding parameters in \cite{bk}. 
}. The counterterm parameters  $\a_i$ are fixed in such a way 
that the {\it renormalized} Euclidean action $I_E$ is finite 
\begin{equation}
I_E\equiv \lim_{r\to \infty}\ \biggl(
S_5^r+S_{GH}+S^{counter}\biggr)\,,\qquad |I_E|<\infty\,.
\eqlabel{IEdef}
\end{equation}
Using the explicit solution \eqref{sol} we find the answer 
\eqref{results} for $I_E$ and 
\begin{equation}
\a_1=\ft 32 g,\qquad \a_2=\ft 18 g^{-1}\,.
\eqlabel{ai}
\end{equation}

We now proceed to the computation of the ADM mass for the 
background \eqref{5action}. 
Following \cite{bk} we define 
\begin{equation}
M=\int_{\Sigma}d^3\xi\ \sqrt{\sigma} N_{\Sigma} \epsilon\,,
\eqlabel{massdef}
\end{equation}
where $\Sigma\equiv S^3$ is a spacelike hypersurface in $\del\calm_5$
with a timelike unit normal $u^\mu$,
$N_{\Sigma}$ is the norm of the timelike Killing vector 
in \eqref{metric5}, ${\sigma}$ is the determinant of the induced metric 
on $\Sigma$, and $\epsilon$ is the proper energy density 
\begin{equation}
\epsilon=u^\mu u^\nu T_{\mu\nu}\,.
\eqlabel{epdef}
\end{equation}
The quasilocal stress tensor $T_{\mu\nu}$ for our background
is obtained from the variation of the full action 
\begin{equation}
S_{tot}=S_5^r+S_{GH}+S^{counter}\,,
\end{equation}
with respect to the boundary metric $\delta h_{\mu\nu}$
\begin{equation}
T^{\mu\nu}=\frac{2}{\sqrt{-h}}\ \frac{\delta S_{tot}}{\delta h_{\mu\nu}}\,.
\eqlabel{qlst}
\end{equation}  
Explicit computation yields 
\begin{equation}
T^{\mu\nu}=\frac{1}{8\pi G_5}\biggl[
-\Theta^{\mu\nu}+\Theta h^{\mu}-2\a_1 h^{\mu\nu}+4\a_2 \left(
R_4^{\mu\nu}-\ft 12 R_4 h^{\mu\nu}\right)\biggr]\,,
\eqlabel{tfin}
\end{equation}
where 
\begin{equation}
\Theta^{\mu\nu}=\ft 12 \left(\nabla^\mu n^\nu+
\nabla^\nu n^\mu\right),\qquad \Theta=\tr \Theta^{\mu\nu}\,.
\eqlabel{thdef}
\end{equation}
It is straightforward to verify that with $\a_i$ as in \eqref{ai},
the mass as defined in \eqref{massdef} is finite, and is given 
by \eqref{results}.

\subsection{The phase diagram of a single charge black hole}
To make the connection with the dual $\caln=4$ $SU(N)$ SYM theory on 
$\reals\times S^3$ we recall \cite{a9905}
\begin{equation}
\frac{1}{g^3 G_5}=\frac{2 N^2}{\pi},\qquad g=\ \ell^{-1} \,.
\eqlabel{g5def}
\end{equation}
We would like to identify the thermodynamics 
characteristics of the single charge black hole 
computed in the previous section $\{I_E,M,S_{BH};T_{H},\mu_{\tq}\}$ with the 
appropriate gauge theory quantities $\{\Omega,E,S,T,\mu_J\}$
\begin{equation}
\{I_E,M,S_{BH};T_{H},\mu_{\tq}\}
\longleftrightarrow
\{\Omega,E,S;T,\mu_J\}\,,
\eqlabel{translation}
\end{equation}
where the thermodynamic potential $\Omega$ is related to the 
Helmholtz free energy $F$ in the standard way
\begin{equation}
\Omega=F-\mu_J J=E-T\ S -\mu_J J\,.
\eqlabel{gmF}
\end{equation}
The identification we are after must preserve the relation
\eqref{thermrel} and satisfy the first law of thermodynamics
for the grand canonical ensemble with $\{T,\mu_J\}$ as independent
variables
\begin{equation}
d\Omega= -S\ dT -J\ d \mu_J\,.
\eqlabel{1st}
\end{equation}
We propose to identify 
\begin{equation}
\begin{split}
T&\equiv T_H\,,\\
S&\equiv S_{BH}\,,\\
\mu_J&\equiv \mu_{\tq}\,.
\end{split}
\eqlabel{initid}
\end{equation}
Given \eqref{initid}, \eqref{thermrel},  the first law 
\eqref{1st} uniquely determines\footnote{Strictly 
speaking there is a single  overall constant, 
independent of temperature and the chemical potential,
in the definition of $E$ and $\Omega$. The latter constant 
must be set to zero in order to reproduce the result for the 
agreement of gauge theory Casimir energy and the 
ADM mass \cite{bk} at $\{T=0,\mu_J=0\}$.} 
\begin{equation}
\begin{split}
\Omega&\equiv T_H\ I_E+\frac{g^2\pi }{12 G_5}\ q^2\,, \\
E&\equiv M+\frac{g^2\pi }{12 G_5}\ q^2\,. 
\end{split}
\eqlabel{finid}
\end{equation}
We do not have an independent way of justifying the 
identification \eqref{finid}, apart from the argument 
presented above. Note that for the vanishing physical 
charge $\tq$, $q$ vanishes as well, and we get the standard 
identification 
$\b F\equiv I_E$. 

In what follows we set the five dimensional gauge coupling 
(or equivalently the $AdS_5$ scale) $g=1$.
Thus, from \eqref{g5def} all the gauge theory thermodynamics quantities 
would scale  $\propto N^2$, as appropriate for the 
deconfined phase. On the other hand, 
in the confined phase we expect these thermodynamic 
potentials to scale $\propto N^0$, effectively zero in the large 
$N$ limit.  
For fixed temperature $T\equiv T_{H}$ and  chemical potential 
$\mu_J$, the physical  gauge theory phase (
in the large $N$ limit\footnote{There is no phase transition, 
but rather a crossover for finite $N$.}) has 
chemical potential $\om_{phys}$ as 
\begin{equation}
\ft {1}{N^2}\ \om_{phys}(T,\mu_J)
=\min\biggl\{\ \ft {1}{N^2}\ \om(T,\mu_J)\,,\ 0\ \biggr\}\,.
\eqlabel{omphys}
\end{equation}
We expect the confinement/deconfinement 
phase transition to occur at $\{T(\mu_J),\mu_J\}$ such that 
\begin{equation}
\om(T(\mu_J),\mu_J)=0\,.
\eqlabel{trans}
\end{equation} 
We found the best way to parametrize the thermodynamics is 
to use the analog of the 'unphysical' charge $q$ in the 
black hole case\footnote{Note that the identification \eqref{initid}
for the chemical potential implies that $J\equiv \tq$.}.  
The summary of the thermodynamics is as follows.
\begin{figure}[ht]
\begin{center}
\epsfig{file=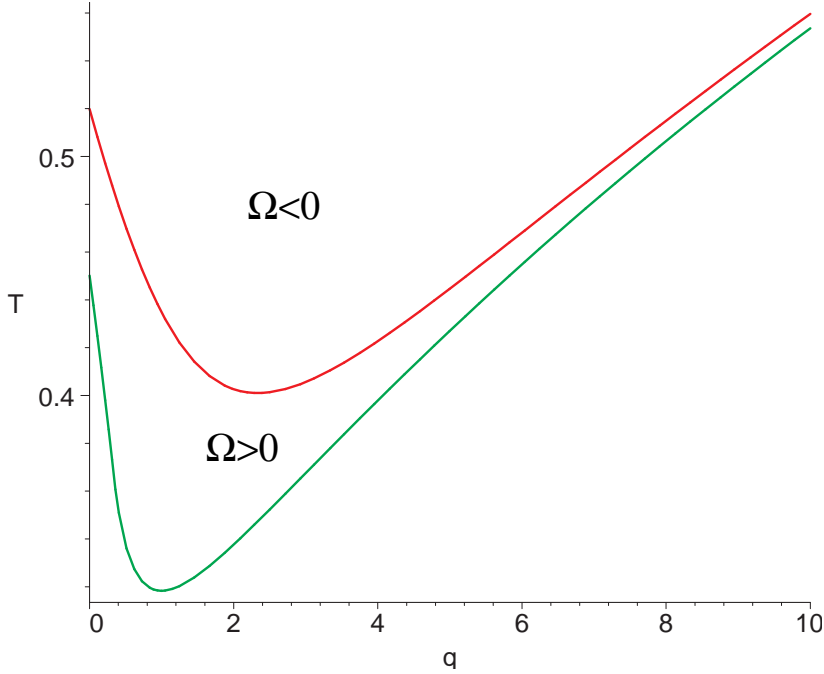,width=0.7\textwidth}
\caption{The critical curves for the thermodynamic potential 
$\Omega(T,q)$, where $q\in[0,+\infty),\ T\in [0,+\infty)$. 
Below the green line $\Im \Omega\ne 0$. 
In a strip between the green and the red lines $\Omega>0$, and 
above the red line $\Omega<0$. The red line corresponds to the 
critical line for the confinement/deconfinement phase transition,
\eqref{trans}.
}
\label{omega}
\end{center}
\end{figure}
\nxt
The critical curves for the generalized free energy $\Omega$ 
as a function of $(T,q)$
are presented in Fig.~\ref{omega}. The green line ( $T_{green}(q)$) 
corresponds to the vanishing of the outer black hole horizon. 
For values of $(T,q)$ below it, both the $\rp$ and the $\Omega$
are imaginary. 
The red line ($T_{red}(q)$) corresponds to the confinement/deconfinement 
phase transition as in \eqref{trans}: for values of $(T,q)$
in the strip between the red and the green lines $\Omega>0$.
For $(T,q)$ above the phase transition curve, $\Omega<0$.   
Asymptotically as $q\to +\infty$ we have
\begin{equation}
\begin{split}
T_{green}&=\ft {1}{2\pi} q^{1/2}+\ft {1}{2\pi} q^{-1/2}\,,\\
T_{red}&=T_{green}+\ft {9}{16\pi} q^{-3/2}+O(q^{-5/2})\,.
\end{split}
\eqlabel{cc}
\end{equation} 
\begin{figure}[ht]
\begin{center}
\epsfig{file=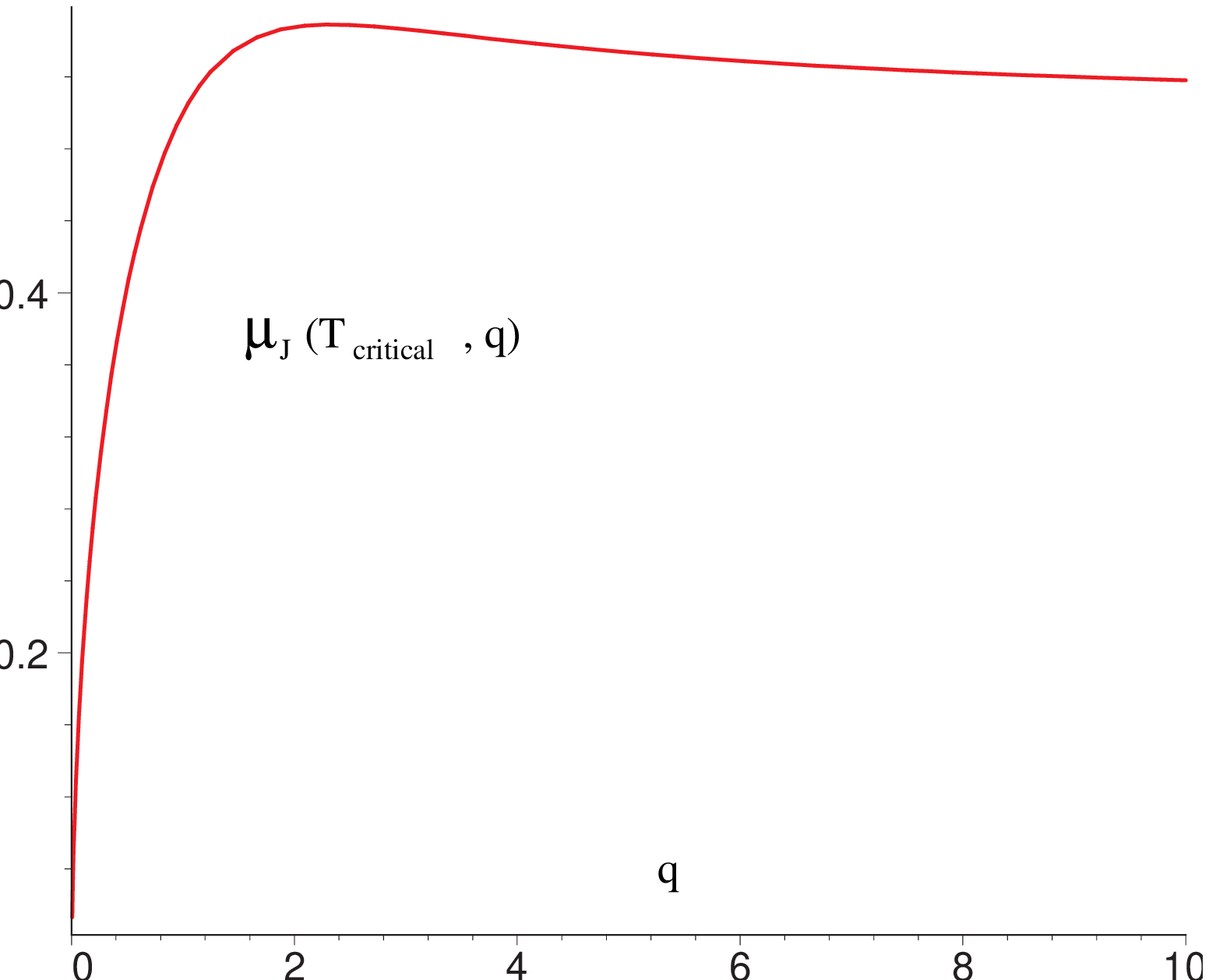,width=0.7\textwidth}
\caption{The chemical potential $\mu_J$ conjugate to the charge $J$ 
as a function of $q$, at the critical temperature $T=T_{critical}$.
}
\label{muJ}
\end{center}
\end{figure}
\begin{figure}[ht]
\begin{center}
\epsfig{file=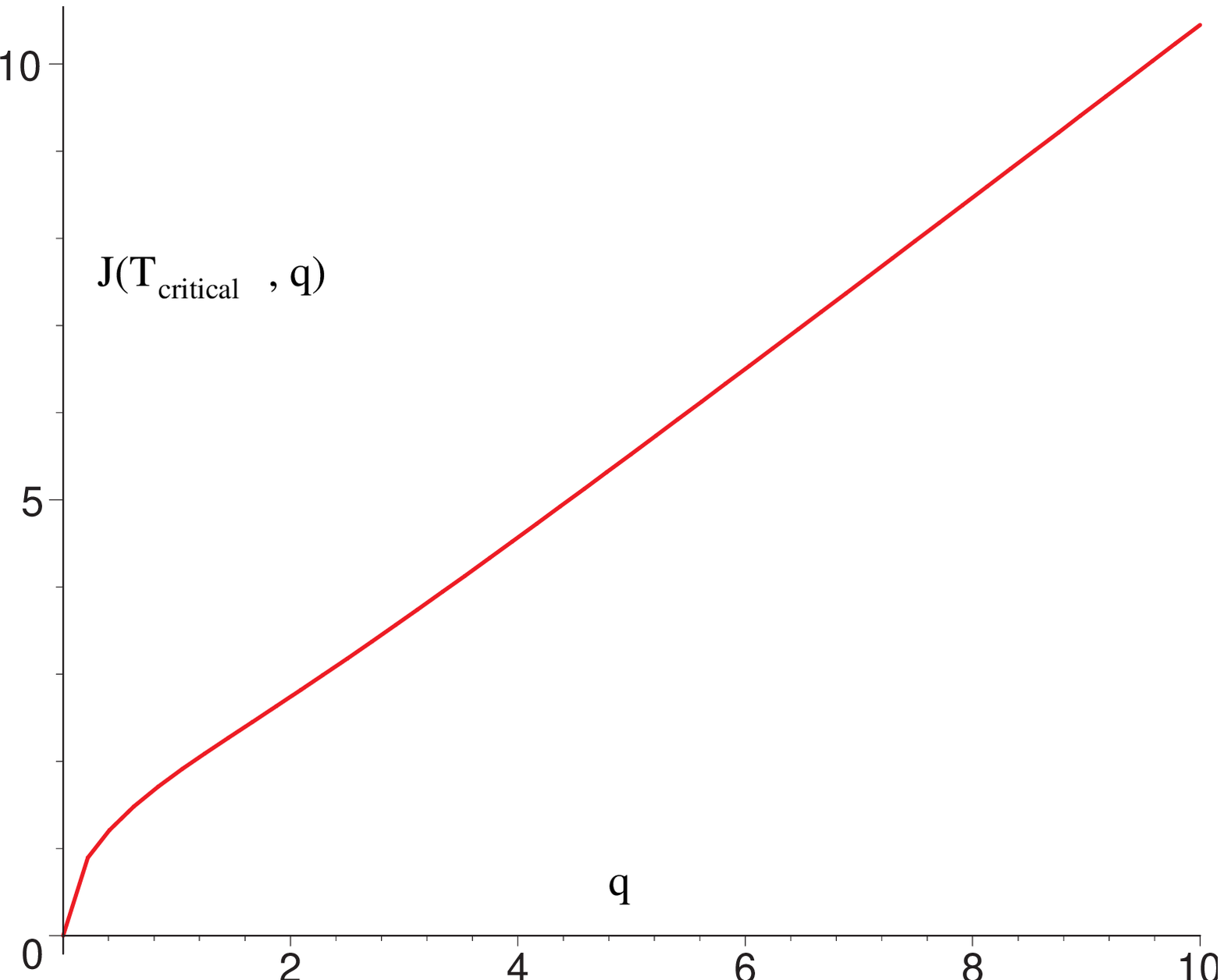,width=0.7\textwidth}
\caption{The $U(1)\subset SO(6)_R$ charge $J$ 
as a function of $q$, at the critical temperature $T=T_{critical}$. 
}
\label{J}
\end{center}
\end{figure}
\nxt
At  the phase transition temperature $T_{critical}\equiv 
T_{red}$, the chemical potential $\ft {\mu_J}{N^2}$ and 
the charge $J$ are shown in Fig.~\ref{muJ} and Fig.~\ref{J}
correspondingly. 
Asymptotically as $q\to +\infty$ we find
\begin{equation}
\frac{\mu_J}{N^2}=\ft 12 +\ft {3}{16} q^{-1} +O(q^{-2})\,,
\eqlabel{muass}
\end{equation}
and
\begin{equation}
J=q+\ft 38 +O(q^{-1})\,.
\eqlabel{jass}
\end{equation}
\nxt
For completeness we also present the asymptotic $q\to +\infty$ 
behavior  of $S_c\equiv S(T_{critical},q)$, $E_c\equiv 
E(T_{critical},q)$, 
$r_{hor}\equiv \rp(T_{critical},q)$ 
\begin{equation}
\begin{split}
\ft {1}{N^2}\ S_c&=
\ft {3\pi}{4} q^{-1/2} +\ft {3\pi}{4} q^{-3/2} +O(q^{-5/2})\,,\\
\ft {1}{N^2}\ E_c&=\ft 12 q +\ft 34 +O(q^{-1})\,,\\
r_{hor}&=\ft {\sqrt{3}}{2} q^{-1/2}+\ft {\sqrt{3}}{4}q^{-3/2} 
+O(q^{-5/2})\,. 
\end{split}
\eqlabel{ser}
\end{equation}

\section{Plane wave background at finite temperature}
One of the most interesting aspects the plane wave background is that,
being a Penrose-G\"uven limit of $AdS_5\times S^5$, it can be used as
an specific example \cite{bmn} of the gauge/string theory  
correspondence on one side of
which there is an exactly solvable theory \cite{metsaev}. 

The aspect we review in this section is the thermodynamic properties
of this background and its implications for the field theory. 
The finite temperature partition function of string theory in the
plane wave background with RR null 5-form field has been 
obtained in \cite{pv,gss,blt,gost}. One of the most salient
features that has been established is the existence of a Hagedorn
temperature for strings in this background.

To clarify the relation between the string theory and the field theory
quantities it is convenient to follow the Penrose limit suggested by
Tseytlin and presented in \cite{bn}. Namely, we start with 
\be
x^+=t, \qquad x^-=R^2(t-\psi),
\ee
where $R$ is the AdS radius and $\psi$ is a coordinate parameterizing the
great circle of $S^5$. The Penrose-G\"uven limit along this null
geodesic results in the standard maximally supersymmetric IIB plane wave
background with null RR 5-form flux and the following relation between
string and gauge quantities:
\be
\label{gauge}
\frac {2p_{+}}{ \m}=E-J, \qquad  2\m \alpha' p_{-}=\frac{J}{
\sqrt{\lambda}}.
\ee
In the field theory interpretation $E$ is the energy of states in
$\reals\times S^3$ and $J$ is the $U(1)$ R-charge of the
corresponding state. 

Let us, following \cite{gss,blt}, define a slightly more general
partition function of the form 
\be
{\cal Z}(a,b;\m)={\rm Tr}_{\cal H}e^{-bp_--ap_+}.
\ee
One of the virtues of this partition function is that it makes
explicit the interpretation in the grand canonical ensemble as
introducing a chemical potential. The partition function of an ideal gas
of IIB strings in the maximally supersymmetric plane wave background can
be written in term of the single string partition function for bosonic
$Z_1^B$ and fermionic $Z_1^F$ modes:
\be
\ln {\cal Z}(a,b;\m)=\sum\limits_{r=1}^\infty \frac{1}{
  r}\bigg[{\cal Z}_1^B(a\,r,b\,r;\m)-(-1)^r {\cal Z}_1^F(a\,r,b\,r;\m)\bigg]. 
\ee
The single string partition function can be written as
\be
\label{single}
{\cal Z}_1(a,b;\m)={\rm Tr}_{\cal H}e^{-bp_--ap_+}.
\ee
Most of our conclusions will rely on the single string approximation. 
Using the expression for the light-cone Hamiltonian obtained in
\cite{metsaev}:
\begin{eqnarray}
H_{lc}&=&\frac{1}{ \a'\,p_-}\bigg[m
\sum\limits_{i=1}^8(a_0^{i\,\dagger}a_0^i+S_0^{i\,\dagger}S_0^i) \nonumber \\
&+&\sum\limits_{n=1}^\infty \sqrt{n^2+m^2}\left(\sum\limits_{i=1}^8
  (a_n^{i\,\dagger}a_n^i
+\tilde{a}_0^{i\,\dagger}\tilde{a}_0^i)+(S_n^{i\,\dagger}S_0^i
+\tilde{S}_0^{i\,\dagger}\tilde{S}_0^i)\right)\bigg],
\end{eqnarray}
where $m=\m\,\a'\,p_-$ and identifying $\tau_2=a/2\pi \,\a' p_-$ we find
\be
Z_1(a,b;\m)=\frac {aV_L}{ 4\pi^2 \a'}\int\limits_0^\infty \frac{d\tau_2}
{
\tau_2^2}\int\limits_{-1/2}^{1/2} d\tau_1 e^{-
\ft {ab}{
2\pi\a'\tau_2}} z_{lc}^{(0,0)}(\tau,\m\,a/2\pi
\tau_2)z_{lc}^{(0,1/2)}(\tau,\m\,a/2\pi \tau_2),
\ee
where $z^{(0,0)}_{lc}$ and 
$z^{(0,1/2)}_{lc}$ are roughly the partition functions of massive
two-dimensional 
bosons and fermions respectively and we refer the reader to \cite{pv}
for the precise notation and further details. 
Our main concern will be with the Hagedorn temperature although other
thermodynamic quantities such as the free energy can also be
calculated \cite{pv,gss, blt,gost}. We define the Hagedorn
temperature as the value above which the partition function starts
diverging. 
\be
\frac {b}{ 16 \m \a'} =\g_0(\frac {a\m}{ 2\pi})-\g_{1/2}(\frac {a\m}{
2\pi}).
\ee 
where on the right hand side figure the difference of the Casimir
energies for bosons and fermions \cite{pv}. The above expression can be
made explicit in terms of integrals or Bessel functions
\cite{pv,gss,blt,gost}.

\begin{figure}[ht]
\begin{center}
\epsfig{file=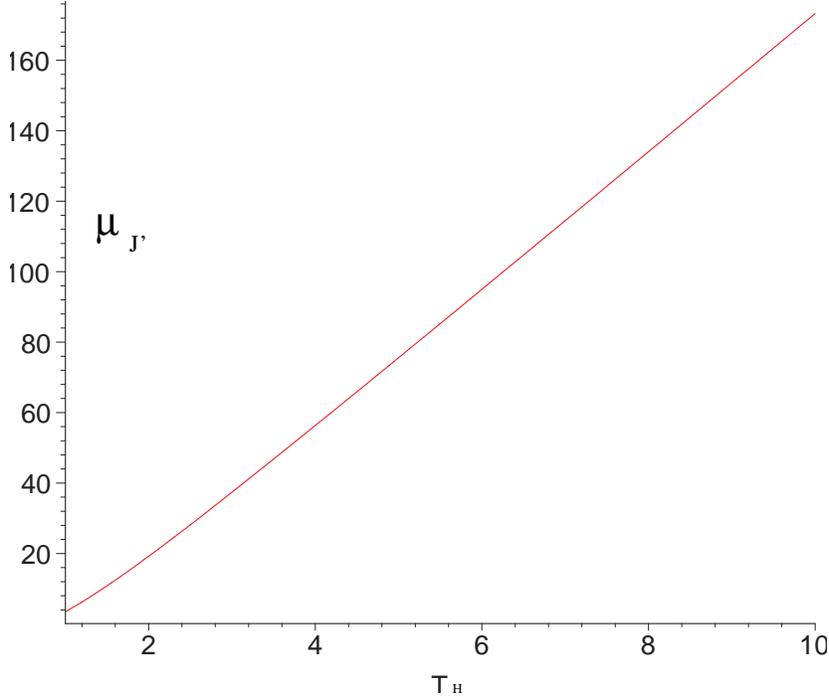,width=0.7\textwidth}
\caption{The chemical potential $\mu_{J'}$ 
as a function of the Hagedorn temperature $T_H$
for strings in the maximally 
supersymmetric plane wave with null RR five form field strength.
}
\label{pp}
\end{center}
\end{figure}

Let us now turn to the meaning of the expression for the Hagedorn
temperature in terms of the field theory variables. 
Taking into account
(\ref{gauge}) and (\ref{single}), the temperature and chemical potential
of the gauge theory are 
\be
\b_{YM}=\frac{1}{ T_{YM}}=\frac{a\m}{ 2}, \qquad \mu_J=\frac{b}{ 2\m  \a'
\sqrt{\la}} -\frac{a\m}{ 2}.
\ee
We will use $b'=b/{\sqrt{\la}}$, or correspondingly introduce 
the chemical potential   conjugate to
$J'=J/\sqrt{\la}$. Note that  the latter quantity  remains finite in the BMN
limit. Thus, the equation for the Hagedorn temperature as a function of
the chemical potential in gauge theory quantities takes the final form:
\be
\frac{1}{ 16}\left(\frac{1}{ T_H}+\mu_{J'}\right)=
\sum\limits_{p=1}^\infty\frac{1}{p}(1-(-1)^p)K_1\left(\frac{2p}{ T_H}\right) \,.
\ee
The dependence of the chemical potential on the Hagedorn temperature 
is presented in Fig.~\ref{pp}.
Notice  that for  high Hagedorn temperatures the relation between the
chemical potential and temperature simplifies to 
\begin{equation}
\mu_{J'}\approx 2\pi^2 T_H\,.
\eqlabel{hag}
\end{equation}

\section{Hagedorn vs. Hawking-Page transition}
The main motivation of this paper is to compare the
Hagedorn behavior of strings in the maximally supersymmetric 
plane wave background, and the phase transition of the 
large charge black holes in the global $AdS_5$ background. 
To reiterate, the hope is that these large charge black
holes are holographically dual to the large R-charge 
sector of the $\caln=4$ SYM at finite temperature, which in turn can be described 
by an exactly soluble string model. The idea is then 
to compare the regime of the phase transition 
of black holes and the Hagedorn transition of strings,
and thus 'unify' the two pictures 
of the confinement/deconfinement phase transition 
in $\caln=4$ SYM on $\reals\times S^3$: 
a Hagedorn behavior at $\la\ll 1$, as discussed 
in \cite{sun}, and the Hawking-Page black hole 
transition at $\la\gg 1$, as discussed in \cite{w97}.
In the rest of this section we explain that 
the regime of the criticality of large charge black
holes appear to be vastly different from the 
regime of the Hagedorn behavior of strings in
PP wave background.

In section 2 we studied the thermodynamics 
of large charge black holes in  global $AdS_5$  geometry.
We found that these nonextremal geometries 
have a phase transition which is, in a sense, a  generalization 
of the Hawking-Page phase transition  \cite{hp}.
It  was argued in \cite{w97} that the HP phase transition  realizes 
the strong coupling gravitational dual to the 
confinement/deconfinement phase transition in $\caln=4$ $SU(N)$
supersymmetric Yang-Mills theory on $\reals\times S^3$ background.
There is an important difference between the phase 
transition for the charged black holes and the HP one discussed in \cite{w97}. 
For the HP transition one has two geometries, and compares the 
difference of their free energies (or Euclidean gravitational action).
In the limit of vanishing nonextremality, one of the geometries does not 
have a horizon, but is simply a global Euclidean $AdS_5$ with the 
appropriately periodically identified time direction. 
In study the phase transition, this geometry is then  
used as a reference for the subtraction 
(``regularization'' in the language of \cite{bk} ) of the 
Euclidean black hole gravitational action.   
As we explained above,  in the case of the charged BH
this subtraction procedure leads to an infinite 
expression for the ADM mass of the BH. Rather, the appropriate 
generalization would seem to be the subtraction of the ``charged''
global $AdS$ geometry, which however does not have a horizon. 
Such a nonsingular geometry does not exist  since it
requires to take the limit  $\mu\to 0$ with $q$ -- fixed  and this is
a violation of the 'BPS bound'. This situation motivated the use
of the regularization procedure, 
which does not require a reference background. 
Luckily, this can be implemented with a straightforward application 
of the holographic renormalization ideas of \cite{bk}. 
The obvious drawback for the absence of the reference 
background is that from the gravity perspective, the phase 
transition is defined in a rather {\it ad hoc} manner, 
\ie, as the vanishing of the generalized free energy,
\eqref{trans}. Nonetheless, this definition of the phase transition 
is well motivated from the perspective of the 
holographically dual gauge theory. 
We found, compare \eqref{cc}-\eqref{jass}, that a 
large value of charge for the black hole at the critical 
(phase transition) temperature implies  a  {\it finite}
nonzero value of the chemical potential.
Also, 
large charge at criticality implies large 
critical temperature:
\begin{equation}
\begin{split}
T_{critical}\bigg|_{J\to \infty}&\propto J^{1/2}\,,\\
\mu_J(T_{critical})\bigg|_{J\to \infty}&\to \frac {N^2}{2}\,.
\end{split}
\eqlabel{last1}
\end{equation}  
In section 3 we reviewed, that quite opposite to 
the regime of the critically of large charge black holes
\eqref{last1}, the  
Hagedorn regime of strings in PP wave background 
for high Hagedorn temperature requires large values 
of the chemical potential, \eqref{hag}.

\section{Conclusion} 
We attempted to provide a more precise relation between the Hagedorn 
description and the Hawking-Page like phase transition 
for the confinement/deconfinement transition in four dimensional 
$\caln=4$ $SU(N)$  theory on $S^3$, by introducing a chemical potential 
conjugate to the large $R$-charge. The hope was that though 
the exact quantization of strings in $AdS_5\times S^5$ is not known,
for large R-charge the essential physics could be captured 
by a string dual to this large R-charge sector, which {\it is}
exactly soluble. 
Unfortunately, we found that this is not so. 
It is plausible that rephrasing this confinement/deconfinement 
phase transition in the language of the Hagedorn transition 
at strong coupling would require the more detailed 
understanding of strings in $AdS_5\times S^5$.

\section*{Acknowledgments}
We are grateful to Finn Larsen, Jim Liu, 
Eliezer Rabinovici and   Cobi Sonnenschein
for useful discussions. LAPZ is especially thankful to D. Vaman for
collaboration in relevant topics. 
This work is supported in 
part by the U.S. Department of Energy.

\end{document}